# Dirac Material Graphene

Elena Sheka

*Russian Peoples' Friendship University of Russia*
*Moscow, 117198 Russia*

sheka@icp.ac.ru

**Abstract.** The paper presents author's view on spin-rooted properties of graphene supported by numerous experimental and calculation evidences. Correlation of odd $p_z$ electrons of the honeycomb lattice meets a strict demand "different orbitals for different spins", which leads to spin polarization of electronic states, on the one hand, and generation of an impressive pool of local spins distributed over the lattice, on the other. These features characterize graphene as a peculiar semimetal with Dirac cone spectrum at particular points of Brillouin zone. However, spin-orbit coupling (SOC), though small but available, supplemented by dynamic SOC caused by electron correlation, transforms graphene-semimetal into graphene-topological insulator (TI). The consequent topological non-triviality and local spins are proposed to discuss such peculiar properties of graphene as high temperature ferromagnetism and outstanding chemical behavior. The connection of these new findings with difficulties met at attempting to convert graphene-TI into semiconductor one is discussed.

## 1. Introduction

Presented in the current paper concerns three main issues laying the foundation of spin features of graphene: (1) a significant correlation of $p_z$ odd electrons, exhibited in terms of spin peculiarities of the unrestricted Hartree-Fock (UHF) solutions, (2) spin-orbital coupling (SOC) and (3) the crucial role of the C=C bond length distribution over the body in exhibiting the spin-based features. Despite numerous and comprehensive studies of graphene performed during the last decade [1], the spin-rooted peculiarities, involved in the graphene physics and chemistry, still remain outside the mainstream. However, graphene is doubtlessly of spin nature and the main goal of the paper is to clarify how deeply electronic states of graphene molecules and solids are spin-burdened.

## 2. Relativistic Electrons of Graphene

Implementing Löwdin's statement "different orbitals for different spins" and revealing correlation effects in electronic spectrum [2], non-relativistic correlation effects in electronic properties are well pronounced and form the ground for peculiarities of chemical [3, 4], magnetic [5], and mechano-chemical [6] behavior of graphene. In contrast, the relativistic SOC contribution related to graphene bodies consisting of light elements is expectedly small (see [7] and references therein). Despite this, the quasi-relativity theory has accompanied graphene crystal from the first consideration of its electronic structure [8, 9]. The primitive cell of graphene crystal is simple and contains two atoms. However, these cells are additionally hexagonally configured to fit the honeycomb lattice, on the one hand, and to provide the hexagonal and flat first Brillouin zone (BZ), on the other. The BZ contains two nonequivalent sets of three vertices K and K' each while Γ point is located at the hexagon centre. In the tight-binding approach it is typical to separate the Hamiltonian for the π (odd $p_z$) electrons from that for the σ electrons, which is strictly valid only for a flat graphene sheet. Thus obtained [10], the total band structure of graphene crystal is of particular image, typical view of which is presented in Fig. 1a. Since referring relativistic analogy concerns π bands, it is conventional to just consider the latter. The relevant low-energy quasiparticle states at the Fermi level, marked by a tinny band in the figure, form six pairs of touching cones with the tips at K (K'), two pairs of which are shown in Fig. 1b. The total low-energy electronic spectrum of the graphene six pairs is described [11] as

$$E_1(k_0 + \kappa) = E^0 - \left(\frac{\hbar p_0}{m}\right)\kappa,$$

$$E_2(k_0 + \kappa) = E^0 + \left(\frac{\hbar p_0}{m}\right)\kappa.$$

(1)

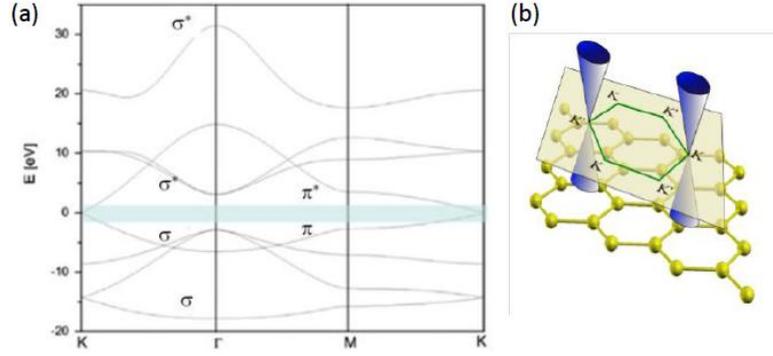

**Figure 1.** (a) Widely presented view on the band structure of graphene. Fermi level is settled at zero. The bands below (above) the Fermi level are related to the valence (conductive) zones. (b) Two pairs of valence/conductive Dirac cones at K and K' points at the Fermi level.

Here $E^0$ and $k_0$ are the Fermi energy and quasiparticle momentum at K (K') points while $E_1$ and $E_2$ spectra are related to the conducting and valence bands, respectively. Detailed description of parameter $\hbar p_0/m$ is given in Ref. 11. Equations (1) are well similar to those related to Dirac's massless fermions due to which the low-energy quasiparticles in the vicinity of K (K') points (Dirac points later) can formally be described by the Dirac-like Hamiltonian

$$\widehat{H} = \hbar v_F \begin{pmatrix} 0 & k_x - ik_y \\ k_x + ik_y & 0 \end{pmatrix} = \hbar v_F \boldsymbol{\sigma} \cdot \boldsymbol{k}, \qquad (2)$$

where $\boldsymbol{k}$ is the quasiparticle momentum, $\boldsymbol{\sigma}$ is the 2D Pauli matrix for pseudospins and the $k$-independent velocity $v_F$ plays the role of the speed of light. The equation is a direct consequence of graphene's crystal symmetry that involves honeycomb hexagonal lattice [9, 11]. Owing to this specific electron band structure, graphene was attributed to the Dirac material and until now graphene has been considered as a 'solid-state toy' for relativistic quantum mechanics [12-14]. Since the graphene massless Dirac fermions move with the Fermi velocity $v_F \sim 10^6$ ms$^{-1}$, it is possible to mimic and observe quantum relativistic phenomena, even those unobservable in high energy physics, on table-top experiments at much lower energies due to small value of the $v_F/c$ ratio. Thus, a quite satisfactory consistence between theoretical predictions and experimental observations has allowed speaking about the observation of Dirac fermions in graphene. Taking them as physical reality, one has suggested a specific engineering of different Dirac fermions by modulating their Fermi velocity when attaching graphene to different substrates [15]. As seen in Fig. 2, an impressive changing of $v_F$ from 1.15 $10^6$ ms$^{-1}$ to 2.46 $10^6$ ms$^{-1}$ is observed when substituting SiC(000$\bar{1}$) substrate by quartz.

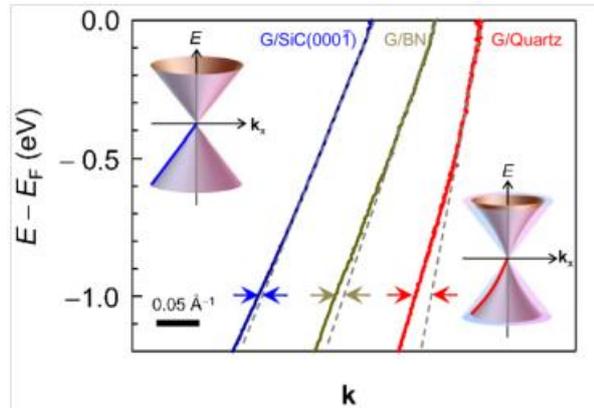

**Figure 2.** Experimental $E_2(k_0 + \kappa)$ dispersions for graphene on SiC(000$\bar{1}$), BN, and quartz. Insets exhibit changing in the graphene Dirac cones when going from weak (left) to strong (right) interaction with substrate [15].

Since Dirac cones are specific crystal symmetry effect, the issue well suits any flat (even quasiflat) hexagonal arrangements of atoms similar to the honeycomb lattice that supply hexagonal BZ. Actually, the Dirac-fermion-like behavior of electronic states were observed for monolayers of silicon atoms on Ag(111) surface voluntarily attributed to 'silicene' species [16] (see detailed discussion of the reality and virtuality of silicene in Ref. 17). Similar behavior was predicted for higher tetrels of group 14 elements - germanene and stanene [18]. Particular attention should be given to a new class of artificial 'molecular graphenes' that mimic honeycomb lattice structure. One of such 'molecule' was synthesized using individually placed carbon monoxide molecules on Cu(111) surface [19]. A completed 'flake' of the molecular graphene is shown in topographic form in Fig. 3a, demonstrating a perfect internal honeycomb lattice and discernible edge effects at the termination boundaries. In spite of finite size of the structure obtained, due to which it should be attributed rather to 'molecular graphene' than to 'graphene crystal', as seen in Fig. 3b, two energy cones are characteristic for the energy band structure near the Fermi level. Estimations showed that the crystal-like behavior well conserves when the molecule size is of 20 *nm* or more.

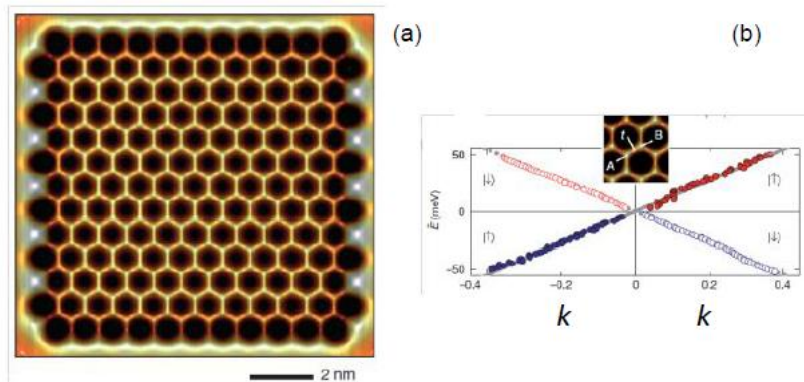

**Figure 3.** (a) Constant current topograph of molecular graphene-like lattice composed of 1549 CO molecules. (b) Linearly dispersing quasi-particles revealed by the conductance spectra, plotted individually for sublattice A (filled circles: pseudospin $s_z$=+1/2) and sublattice B (open circles: pseudospin $s_z$=-1/2), measured at locations *t* illustrated in the inset. Adapted from Ref. 19.

The other quite peculiar ability to create artificial graphene-like structure utilizes an optical honeycomb lattice to trap ultracold potassium atoms [20]. Dirac-cone-like band structure is reproduced in this system as well. This optical method of creating the honeycomb lattice suggests large possibility to investigate factors influencing the Dirac cones structure. Thus, by tuning the anisotropy of the lattice, the locations of the Dirac points may be shifted. When the anisotropy reaches a certain limit, the Dirac points merge and annihilate, while the evidence supporting the existence of a theoretically predicted topological phase transition was observed.

A number of theoretical suggestions on the Dirac-graphene-like structure is quite impressive. It covers virtual silicene, germanene, stanene (see review [21] and references therein), hydrogenated borophene [22] and arsenene [23]. All the Dirac species are described by hexagon packing of two-atom primitive cells. However, 'the primitive cell' may be considerably complicated as it takes place in the case of s-tirazines with primitive cells composed of either $C_6N_6$ or $C_{12}N_6$, and $C_{24}N_6H_{12}$ molecular compositions [24], graphitic carbon nitride (GCN) with $C_{14}N_{10}$ as a primitive cell [25], beautiful hexagon patterned lace of $NiC_8S_2H_4$ molecules [26], the FeB2 monolayer with graphene-like boron sheet [27], an impressive number of MXenes [28] (a new class of inorganic 2D compounds [29]), just appeared new compounds InSe [30] and so forth. The conservation of the hexagon packing of primitive cells mentioned above protects the presence of Dirac cones in the electronic spectra of all the species.

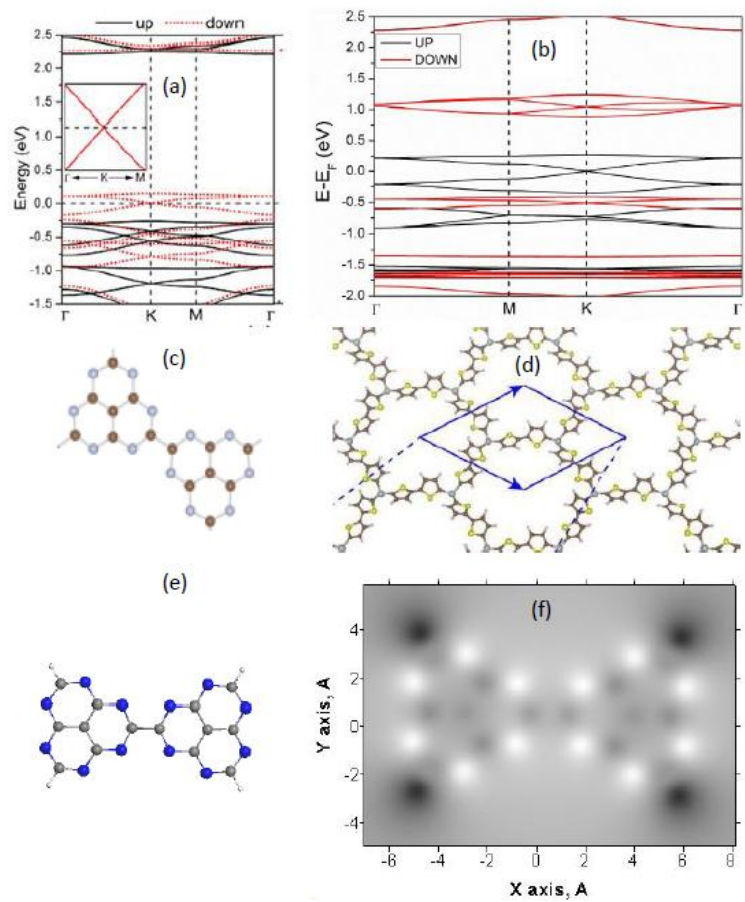

**Figure 4**. Spin-polarized band structure and primitive cell of GCN $C_{14}N_{10}$ (a) and (c) and $Ni_2C_{24}S_6H_{12}$ (b) and (d). Adapted from Refs. 25 and 26, respectively. Equilibrium structure (e) and the ACS $N_{DA}$ image map of local spin distribution (f) for the $C_{14}N_{10}H_4$ molecule; UHF AM1 calculations.

Virtually all the Dirac spectra discussed above were calculated not paying attention to if the studied system is open- or closed-shell one and exploiting closed-shell formalism. Only calculations related to GCN $C_{14}N_{10}$ [25] and metal-organic framework with primitive cell

Ni$_2$C$_{24}$S$_6$H$_{12}$ [26] were obtained taking into account that electrons with $\alpha$ and $\beta$ spins are correlated and separated in space. The open-shell approach immediately revealed spin-polarization of the electronic spectra just doubling the band number and combining them in $\alpha$ and $\beta$ sets. Figure 4a presents the spin-polarized band structure of GCN while Fig. 4b is related to metal-organic framework. The configurations of the relevant primitive cells are shown under the spectra. Both primitive cells contain even number of valence electrons, $N^\alpha = N^\beta$ so that in the two cases there are no unpairing free spins since total spin density is zero. In the case of GCN, the authors [25] explain the observed spin polarization from a chemical bonding analysis and attributed it to reducing the anti-bonding characteristics and density of states at Fermi level.

However, it is quite reasonable to suggest an alternative explanation and connect the obtained spin polarization caused by $p_z$ electron correlation with open-shell character of the electronic system of both cells [7]. Actually, as shown in Fig. 4e, in view of the UHF formalism the molecule C$_{14}$N$_{10}$H$_4$, which perfectly mimics the GCN primitive cell, is open-shell one with the total number of effectively unpaired electrons $N_D$ = 5.34 $e$ that are distributed as 'local spins' over nitrogen and carbon atoms with average $N_{DA}$ fractions of 0.285 ± 0.001 $e$ and 0.145 ± 0.003 $e$, respectively (see Fig. 4f). Therefore, the band spin polarization becomes the manifestation of the electron correlation in 2D open-shell solids. As seen in Fig. 4, the spin polarization is well pronounced through over both BZs while at K points the Dirac spectra still remain gapless. Apparently, the feature is still caused by the hexagon packing of the molecules. Opening the gap at Dirac points needs the participation of SOC that has been so far ignored.

Disclosed connection between the open-shell character of the electronic system and spin polarization of the electronic band spectra brings us back to graphene. First the character peculiarity was mentioned at addressing graphene magnetism and was illustrated by a convincing spin density map related to the rectangle graphene molecule (5, 5) NGr with five benzenoid units along the armchair and zigzag edges, respectively [5]. Second it concerned the dispersion of the C=C bond length related to the body [31]. Now there is the third time when we have to realize the inevitability of spin polarization of the graphene electronic band structure while the relevant results concerning the band structure of graphene are still in future while Fig. 5 presents the energy splitting related to a selected number of spinorbitals of the (5, 5) NGr molecule. A comparison of RHF and UHF results allows exhibiting correlation effects related to the studied open-shell molecule. The degeneracy of the RHF solution is caused by both high spatial ($D_{2h}$) and spin symmetry of the molecule. As seen in the figure, when going from RHF to UHF formalism the UHF orbitals become clearly split into two families related to α and β spins. The splitting value is different for different orbitals ranging from 0.02 $eV$ to 1.12 $eV$. The orbitals splitting exhibits breaking spin symmetry that causes a remarkable distinguishing of orbitals related to different spins.

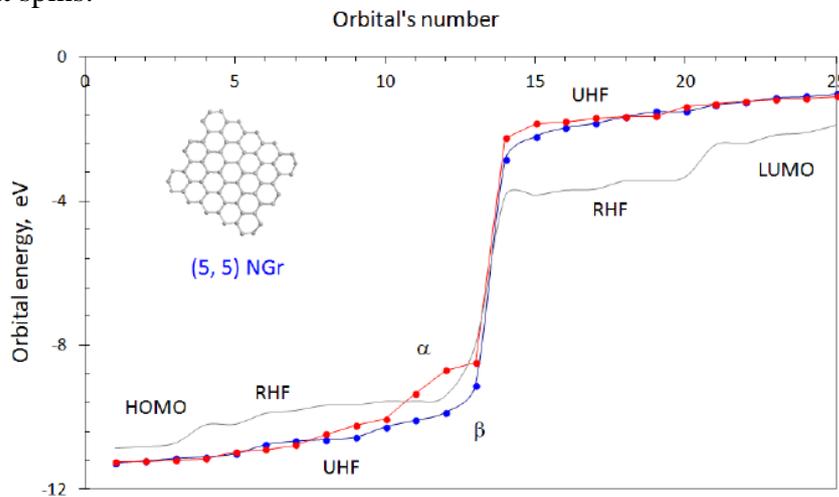

**Figure 5.** Energies of 25 spinorbital in the vicinity of HOMO-LUMO gap of (5, 5) NGr molecule with bare edges; UHF AM1 calculations.

As for the Dirac spectrum shown in Fig.1, as was mentioned, spin polarization does not affect the cones touching (see insert in Fig.4a). However, the electron correlation generates a dynamical SOC (see Ref. 32 and references therein) so that it might be possible to expect some relativistic features, observable at even negligible intrinsic SOC but at a significant electron correlation. The first potential effect concerns the splitting of the Dirac spectrum. To examine if the splitting can be observed for graphene experimentally, the team of Novoselov and Geim performed a particular investigation of how close one can approach the Dirac point [33]. It was shown that the approach value $\delta E$ depends on quality and homogeneity of samples, on the resolution of experimental equipment, on temperature, and so forth. The best value $\delta E$ related to free standing sample constitutes ~1 *meV* at 4 K, thus establishing that there is no bandgap in graphene larger than 0.5 *meV* and that a combined SOC effect is less this value. Nevertheless, the finding does not disprove the existence of SOC as such, which may be important in the case of other effects, more sensitive to weak SOC. One of such potential effects concerns the topological non-triviality of graphene.

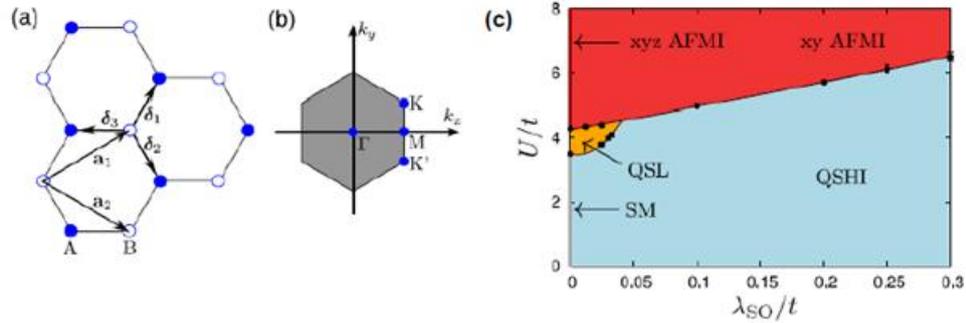

**Figure 6** (a) Kane-Mele model of the honeycomb lattice consisted of two sublattices A, B. (b) The hexagonal first Brillouin zone contains two nonequivalent Dirac points K and K`. (c) Phase diagram of the half-filled Kane-Mele-Hubbard model from quantum Monte Carlo simulations. Adapted from Ref. 32.

Actually, electron correlation and SOC are crucial characteristics of the topological non-triviality of 2D bodies. This question has been clear since the very time of the Dirac topological insulator (TI) discovery [34]. In the case of graphene, negligible SOC and complete ignorance of the electron correlation were major obstacles to a serious discussion of this issue. However, the topological non-triviality covers a large spectrum of different topological states and phases involving both ideal Dirac TIs (or quantum spin Hall insulators – QSHI) and other topological issues such as correlated topological band insulators, interaction-driven phase transitions, topological Mott insulators and fractional topological states [32] interrelation between which is determined by that one between SOC and correlation effects. Figure 6 presents a phase diagram of topological states characteristic for 2D graphene-like honeycomb lattice (known as Kane-Mele model [9]) in relative coordinates of correlation energy ($U$) and SOC ($\lambda_{SO}$), which correspond to the Hubbard model. As seen in the figure, in the limit case $\lambda_{SO}$=0, the relevant 2D structures should be attributed to either semimetal (SM) or antiferromagnetic Mott insulator (AFMI) with Heisenberg order (xyz), depending on the correlation energy. The SOC increasing transforms SM into QSHI at a rather large scale of the correlation energy variation. When the value achieves the critical one shown by the straight line, the QSHI transforms into AFMI with easy plane order (xy). In the limit case $U$=0, the SM should behave as QSHI at all $\lambda_{SO}$.

Concerning graphene, recent estimation of $U/t = 1.6$ [35] allows placing graphene far below the border with the AFM phases and attributing it to the SM phase if $\lambda_{SO} = 0$. However, the doubtless presence of the correlation of graphene $p_z$ electrons causes breaking of Kramers pairs of spinors [36], which violates the time-reversal symmetry, on the one hand, and stimulates the origin of dynamic SOC, on the other [32]. The findings shift graphene along the $\lambda_{SO}/t$ axis in the depth of the QSHI phase thus providing a vivid topological non-triviality of graphene that might be revealed by not only the SOC-stimulated energy gap splitting. One of such topological effects has a direct bearing to peculiarities of graphene magnetism.

## 3. Molecular Essence and Topological Character of Graphene Magnetism

### 3.1. General Features of Experimental Observations

Repeatedly controlled extended graphene sheets are diamagnetic and magnetic response from large samples was empirically obtained only after either, say, heavy irradiation by proton beams or chemical modification (hydrogenation, oxidation and so forth) of graphite and/or graphene (see [37] and references therein). Thorough analysis, performed in each case, allowed excluding the impurity origin of these features and attributing them to graphite/graphene itself, albeit induced by either micro- and/or nanosructuring of samples or by defects of different topology created in due course of chemical modification (see some examples [38-41]). It is important to mention that practically in all the cases a ferromagnetic response at room temperature was observed for graphene species with zero total spin density.

Another scenario concerns magnetic graphene of a paramagnetic behavior [40-42] recorded after either fluorination or bombarding graphene laminates consisting of 10-50 *nm* sheets by protons. The treatment provided the rupture of double C=C bonds inducing 'spin-half paramagnetism' in graphene. In both cases, the magnetization is weak and is characterized by one spin per approximately 1,000 carbon atoms. The ratio indicates that, actually, the after-treatment magnetic crystal structure differs from the pristine one and the difference concerns the unit cell that becomes ~33/2 times larger than the previous one. Besides, the unit cell contains one additional spin thus lifting the spin multiplicity to doublet. Therefore, introduced adatoms and point defects cause a magnetic nanostructuring of the pristine crystal but with non zero spin density.

A doubtless confirmation that nanostructuring of graphene sheets plays a governing role in magnetization was obtained in the course of specifically configured experiments [43, 44]. In both cases the matter is about meshed graphene or graphene in pores that were formed by either fully hydrogenated (oxidized) graphene discs in the first case or MgO nanoparticles in the second. A view on the sample and obtained results related to the first case can be obtained by looking at Fig. 7. A large graphene sheet is put on porous alumina template (see Fig. 7a). The sample was subjected to either one-side hydrogenation or oxidation through alumina pores thus leaving graphene webs between the pores untouched. The web width *W* in a set of alumina templates differed from 10 to 50 *nm*. Ferromagnetic response of the webs at room temperature is presented in Figs. 7b and 7c at different web widths and inset in Fig. 7c discloses the width dependence more clearly. In the second case, chemically modified dark disks in Fig. 7a were substituted with MgO nanoparticles a set of which was covered by CVD grown graphene tissue [44]. The web width between the particles constituted ~ 10 *nm*. The magnetic response from the sample is similar to that presented in Fig.7b while the signal from much larger pieces ~100 *nm* of technical graphene (reduced graphene oxide) was a few time less.

Therefore, nanosize occurrence and size-dependence, on the one hand, and high-temperature ferromagnetic character, on the other, are two peculiar features of zero-spin-density graphene magnetism. Evidently, the former concerns the magnetization magnitude and is associated with molecular essence of graphene while the latter is relevant to the magnetism grounds and applies to its physics thus revealing the molecular-crystalline dualism of graphene once more.

### 3.2. Magnetic Behavior of Graphene Molecules

Zero-spin-density graphene implies the absence of free spins since graphene belongs to species for which $N^\alpha = N^\beta$. Usually such magnetic species were attributed to 'singlet magnets' (in terms of closed-shell approximation) and their magnetization was associated with the effect of the second order of perturbation theory (PT) implying the mixture of the ground singlet state with higher laying states of higher spin multiplicity. If the mixture is caused by the application of magnetic field, the effect is known as van Fleck magnetization [45]. However, graphene is open-shell species due to which its singlet ground state has already been spin-mixed due to odd $p_z$ electron correlation. Since the spin contamination is the PT second order effect as well [7], there is no need to apply to van Fleck effect for the magnetism explanation. The magnetization ability of graphene has been already ensured by its electronic system.

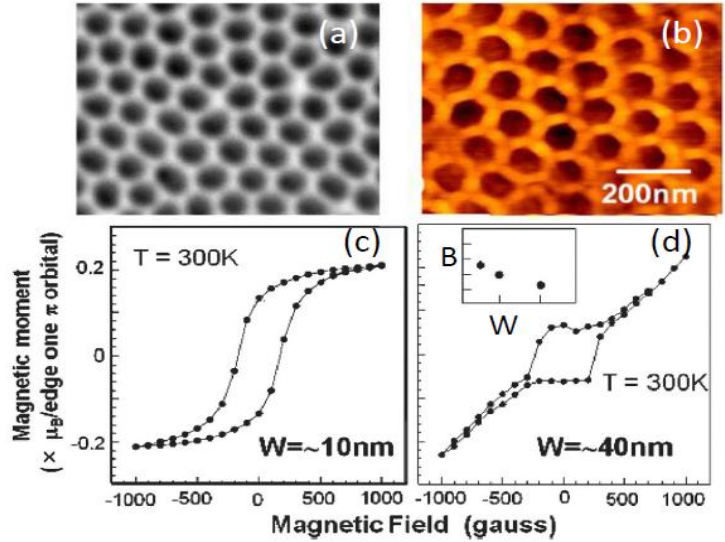

**Figure 7.** (a) SEM image of nanoporous alumina template with mean pore diameter ~80 nm and mean interpore spacing W ~20 *nm*. (b) AFM image of graphene nanopore array formed by using (a) as an etching mask for the sample one-side hydrogenation. (c) and (d) Magnetization of monolayer graphene nanopore arrays at W ~10 *nm* and 40 *nm*. Inset presents the dependence of the residual magnetization of graphene webs on their width W. Adapted from Ref. 43.

The observation of the PT second order contributions strongly depends on the energy denominator among other factors. For covalent species triplet states are the main contributors due to which the energy denominator is $2|J|$, where $J$ is the exchange integral that determines the energetic dependence of pure spin states in terms of the Heisenberg Hamiltonian $H_{ex} = JS(S + 1)$. The integral is usually referred to as magnetic coupling constant [46]. A correct computation of the constant is quite complicated. Happily, about four decades ago Noodelman suggested a simple expression for the value determination for open-shell molecules in the framework of the broken spin symmetry approximation [46]

$$J = \frac{E^U(0) - E^U(S_{max})}{S_{max}^2}. \tag{3}$$

Here, $E^U(0)$ and $E^U(S_{max})$ are energies of the UHF singlet and the highest-spin-multiplicity states, the latter corresponding to the $S_{max}$- pure-spin state. Thus obtained value is widely used and once attributed to molecular magnetism showed [48] that measurable magnetization response can be fixed if $|J| \leq |J_{crit}|$ where empirically estimated $J_{crit}$ is $10^{-2}$ -$10^{-3}$ *kcal/mol*. Basing on the molecular essence of graphene magnetism, let us look which $J$ values can be expected for graphene molecules.

Table 1 lists sets of three criterial quantities: $\Delta E^{RU}$, $\Delta \langle S \rangle^2$, and $N_D$, which characterize any open-shell molecule [4]. The data were evaluated for a number of graphene molecules presented by rectangular ($n_a$,$n_z$) fragments ($n_a$ and $n_z$ count the benzenoid units along armchair and zigzag edges of the fragment, respectively). Consequently, the table as a whole presents the size dependence of the UHF peculiarities of the open-shell graphene molecules. As seen in the table, the parameters are certainly not zero, obviously greatly depending on the fragment size while their relative values are practically non size-dependent. The attention should be called to rather large $N_D$ values, both absolute and relative, that manifest the measure of the $p_z$ odd electrons correlation. It should be added as well that the relation $N_D = 2\Delta \hat{S}^2$, which is characteristic for spin contaminated solutions in the singlet state, is rigidly kept over all the molecules. The data are added by the magnetic constant $J$ determined following Eq. 3.

**Table 1.** Identifying parameters of the odd electron correlation in rectangular nanographenes

| $(n_a, n_z)$ NGrs | Odd electrons $N_{odd}$ | $\Delta E^{RU}$ [1] kcal/mol | $\delta E^{RU}$ %[2] | $N_D$, e | $\delta N_D$, %[2] | $\Delta \hat{S}_U^2$ | J, kcal/mol |
|---|---|---|---|---|---|---|---|
| (5, 5) | 88 | 307 | 17 | 31 | 35 | 15.5 | -1.429 |
| (7, 7) | 150 | 376 | 15 | 52.6 | 35 | 26.3 | -0.888 |
| (9, 9) | 228 | 641 | 19 | 76.2 | 35 | 38.1 | -0.600 |
| (11, 10) | 296 | 760 | 19 | 94.5 | 32 | 47.24 | -0.483 |
| (11, 12) | 346 | 901 | 20 | 107.4 | 31 | 53.7 | -0.406 |
| (15, 12) | 456 | 1038 | 19 | 139 | 31 | 69.5 | -0.324 |

[1] AM1 version of UHF codes of CLUSTER-Z1 [49]. Presented energy values are rounded off to an integer
[2] The percentage values are related to $\delta E^{RU} = \Delta E^{RU} / E^R(0)$ and $\delta N_D = N_D / N_{odd}$, respectively

As seen in Table 1, quantities $\Delta E^{RU}$, $N_D$, and $\Delta \langle S \rangle^2$ gradually increase when the size grows due to increasing the number of atoms involved in the pristine structures. The relative parameters $\delta E^{RU}$ and $\delta N_D$ only slightly depend on the NGr size just exhibiting a steady growth of the parameters. In contrast, $J$ values demonstrate quite different behavior. They show a significant size-dependence, gradually decreasing by the absolute magnitude when the size grows. This dependence can be obviously interpreted as the indication of strengthening the electron correlation thus exhibiting the collective character of the event. The finding is expected to lay the foundation of peculiar size-effects for properties that are governed by these parameters, first of which can be addressed to molecular ferrodiamagnetism. The diamagnetic behavior is provided by σ electrons while the ferromagnetic contribution is obviously related to odd $p_z$ ones.

As mentioned earlier, the primitive cell of graphene crystal, which determines magnetic properties of ideal crystal, involves two atoms joint by one C=C bond of a benzenoid unit. Estimation of $|J|$ value for ethylene and benzene molecule with stretched C=C bonds up to 1.42Å in length gives $|J|$ values of 13 *kcal/mol* and 16 *kcal/mol*, respectively. Despite ethylene and benzene molecules do not reproduce the unit cell exactly a similar $|J|$ constant of the cell is undoubted. Owing to this the crystal should demonstrate the diamagnetic behavior. To provide a remarkable 'ferrodiamagnetism' means to drastically decrease the magnetic constant $|J|$. While it is impossible for regular crystal, graphene molecules are more labile. Shown in Table 1, the least $|J|$ of 0.3 kcal/mol is still large to provide a recordable magnetization of (15, 12) NGr molecular magnet, but the tendency is quite optimistic. Supposing the quantity to be inversely proportional to the number of odd electrons, it is possible to estimate the electron number which would satisfy $|J_{crit}|$ of $10^{-2}$-$10^{-3}$ kcal/mol which gives us $N \sim 10^5$ e.

For rectangular NGrs $N$ odd electrons are supplied by N carbon atoms that, according to [50], is determined as

$$N = 2(n_a n_z + n_a + n_z) \qquad (4)$$

To fit the needed N value, the indices $n_a$ and $n_z$ should be of hundreds, which leads to linear sizes of the NGrs from a few units to tens *nm*. The estimation is rather approximate, but it, nevertheless, correlates well with experimental observations of the ferromagnetism of activated carbon fibers consisting of nanographite domains of ~2 *nm* in size [51] as well as with the data related to meshed graphene [43, 44] discussed earlier. The maximum effect was observed at the interpore distance of 20 nm [43] after which the signal gradually decreased when the width increased. The behavior is similar to that obtained for fullerene oligomers [52], which led to the suggestion of a scaly mechanism of nanostructured solid state magnetism of the polymerized fullerene $C_{60}$ that was confirmed experimentally.

The said above highlights another noteworthy aspect of the graphene magnetism attributing the phenomenon to size-dependent ones. The latter means that the graphene magnetization is observed for nanosize samples only, moreover, for samples whose linear dimensions fit a definite interval, while the phenomenon does not take place at either smaller or bigger samples outside the critical region. Actually, an individual benzenoid unit (and benzene molecule) is only diamagnetic. When the units are joined to form a graphene-like benzenoid cluster, effectively unpaired electrons appear due to weakening the interaction between $p_z$ odd electrons followed by stretching C=C bonds which causes these electrons correlation. The correlation accelerates when the cluster size increases, which is followed with the magnetic constant $|J|$ decreasing until the latter achieves a critical value that provides a noticeable fixation of the spin mixing of the cluster ground state. Until the enlargement of the cluster size does not violate a molecular (cluster-like) behavior of odd electrons, the sample magnetization will grow. However, as soon as the electron behavior becomes spatially quantized, the molecular character of the magnetization will be broken and will be substituted by that one determined by the electron properties of the primitive cell. Critical size parameters, controlling quantization of molecular properties obviously depends on the kind of quasiparticles to be considered. Addressing graphene magnetization, evidently it is Dirac fermions that control the quantizing and their mean free path $l_{fm}$ determines the critical size parameter: when the cluster size exceeds $l_{fm}$ the spatial quantization quenches the cluster magnetization.

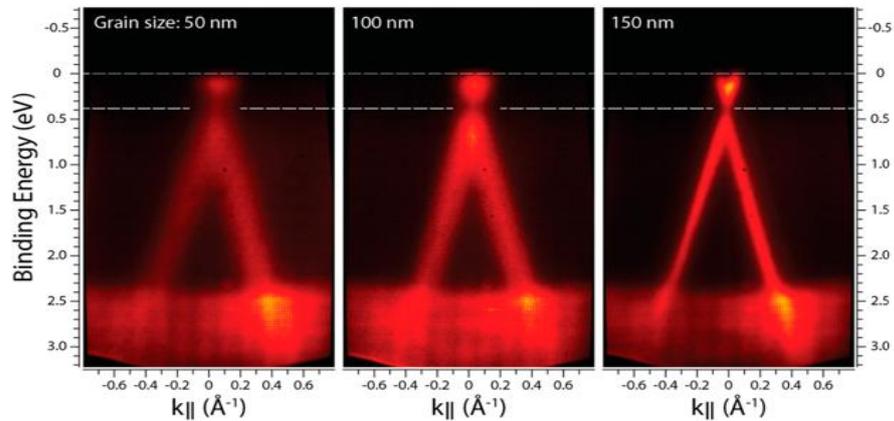

**Figure 8.** Dirac's cones of continuous graphene film with average grain sizes of 50, 100, and 150 *nm* at the K point of graphene Brillouin zone obtained by ARPES mapping. Fermi energy is settled to zero. Adapted from Ref. 53.

Happily, just recently experimental data related to the study of size dependence of both the linearity of the fermion low-energy band $E_{fm}(k)$ within the Dirac cones in the vicinity of the Fermi level and the shape of the spectrum were published. Figure 8 presents a set of $E_{fm}(k)$ spectra related to a polycrystalline graphene sample consisting of grains different in size [53]. As seen in the figure, quantizing is well supported in grains of 150 *nm*, starts to be distorted in grains of 100 *nm* and is remarkably violated for grains of 50 *nm*. A considerable broadening of the spectrum in the last case allows putting the upper bound for $l_{fm}$ around 50 *nm*. A comparable $l_{fm}$ value of ~20 *nm* follows from the data related to CO-hexagon structure [19].

Obviously, the transition from localized to quantized state is not abrupt. Thus at the pore width $W = 40$ *nm* the residual magnetization only halves the maximum value at 20 *nm* (see inset in Fig. 7c) and continuous approaching zero may cover quite a large pore width. Actually, in the case of MgO pores [44], magnetization of rGO flakes with width ~100 *nm* constitutes ~20% of the value at the pore width of 10 *nm*. Nevertheless, the molecules linear size is evidently the governing factor for the magnitude of ferrodiamagnetism of graphene molecules.

### 3.3. High-Temperature Ferromagnetic Topological Insulating Phase of Graphene

If discussed in the previous section allows understanding when magnetic behavior of graphene becomes measurable, it does not answer question why the behavior is ferromagnetic and still exists at room and higher temperatures. Actually, it is difficult to expect ferromagnetism from the species with zero total spin density in the ground state. Additionally, molecular magnetism is usually observed at quite low temperatures [48] and its fixation at room temperature looks highly unexpected. At the same time there are physical objects for which high-temperature ferromagnetism is a characteristic mark. Thus, we come again to peculiar Dirac materials known as TIs [34]. As shown in Section 2, quite considerable electron correlation and small, but available, intrinsic-dynamic SOC allow attributing graphene to weak QSHI. Evidently, the topological non-triviality is relevant to both crystalline and molecular graphene. Accepting this idea shows one of the ways towards the explanation of high-temperature ferromagnetism of graphene.

In view of electron correlation, graphene presents a honeycomb structure of carbon atoms with local spins distributed over them. The spin values are not integer and are determined by the relevant $N_{DA}$ map similar to that presented in Fig. 4f. Evidently, the exchange interaction between the local spins is responsible for the magnetic behavior of graphene. To determine the type of the behavior, let us use the formalism suggested for describing the magnetic impurities on the surface of a topological insulator [54]. In the presence of magnetic impurity or local spins, the main Hamiltonian, describing the TI band structure in the form of Eq. 2, is substituted by new one

$$H = h v_F (\boldsymbol{k} \times \hat{\boldsymbol{z}}) \cdot \boldsymbol{\sigma} - H_{ex} \tag{5}$$

where $v_F$ is the Fermi velocity, $\hat{\boldsymbol{z}}$ is the surface unit normal, $\boldsymbol{\sigma}$ is the Dirac electron spin and

$$H_{ex} = \sum_r J_z s_z(\boldsymbol{r}) S_z(\boldsymbol{r}) + J_{xy}(s_x S_x + s_y S_y) \tag{6}$$

Here $S_i(\boldsymbol{r})$ is the spin of a magnetic impurity located at $\boldsymbol{r}$, $s_i(\boldsymbol{r}) = \psi^*(\boldsymbol{r}) \sigma^i \psi(\boldsymbol{r})$ is the spin of the surface electrons and $J_z$ and $J_{xy}$ are the coupling parameters. When the impurity spin is polarized in z direction the second term in Eq. 6 disappears. As every magnetic impurity will open a local gap in its vicinity, one may expect the system to be gapped everywhere, at least in the mean-field level. However, this is not necessarily true if the magnetization of magnetic impurities is non-uniform. Meeting the problem and comparing the formation of magnetic domain wall and ferromagnetic arrangement, the authors [54] came to the conclusion that magnetic impurities must be ferromagnetically coupled.

Sharing this viewpoint, a similar Hamiltonian $H_{ex}$ was suggested to describe the Dirac-fermion-mediated ferromagnetism in a topological insulator [55]. The Hamiltonian $H_{ex}$ reads

$$H_{ex} = J n_s \bar{S}_z \sigma_z \tag{7}$$

Here $\sigma_z$ is the z component of the electron spin and $n_s$ is the areal density of localized spins with average z component $\bar{S}_z$. $J$ describes the exchange coupling between the z components of the Dirac electron spin $\boldsymbol{\sigma}$ and the local spin **S**, locking $\boldsymbol{\sigma}$ perpendicular to the momentum **k**. Following the same conclusion that every local spin opens the gap and the system must be gapped everywhere one has to accept the necessity of a ferromagnetic configuration for local spins. Apparently, it is the consequence that explains ferromagnetic behavior of pure graphene samples.

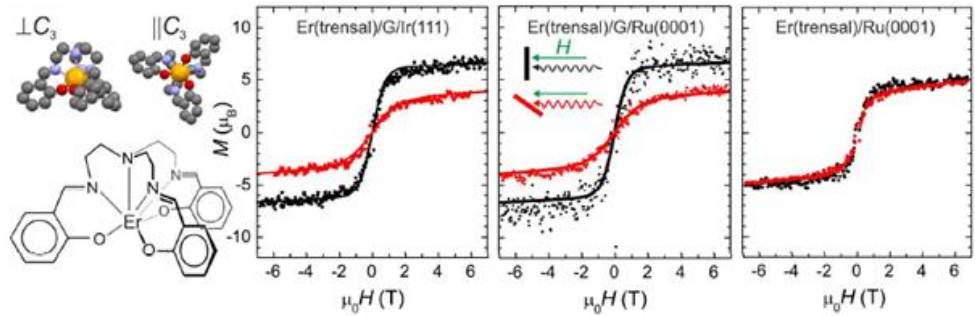

**Figure 9.** Structure and scheme of the molecular magnet Er(trensal). Coloring: orange: Er; blue: N; red: O; grey: C; H atoms are omitted for clarity. From left to right: magnetization curves at normal (black) and grazing (red) orientation of magnetic field with respect to the substrate surface. Adapted from Ref. 56.

Highly convincing evidence, strongly supporting suggestion that graphene is a typical TI, was received in the most recent [56]. Figure 9 presents the accumulation of the main results of the performed study. A molecular complex, presented by Er(trensal) single-ion magnets, was adsorbed on graphene/Ru(0001), on graphene/Ir(111) and on bare Ru(0001) substrates. On graphene, the molecules self-assemble into dense and well-ordered islands with their magnetic easy axes perpendicular to the surface. In contrast, on bare Ru(0001) the molecules are disordered exhibiting only weak directional preference of the easy magnetization axis. Accordingly, the ferromagnetic response is spin polarized in the two former cases while unpolarized in the case of Ru(0001) substrate and additionally twice less by magnitude. Therefore, topologically trivial bare ruthenium surface has no effect on the molecular impurity ordering while the addition of a monolayer graphene covering leads to ferromagnetic ordering of the impurity characteristic for topologically non-trivial substrates, which were discussed above. Not only ordering but enhancement of ferromagnetic response evidences the TI nature of the graphene component of the hybrid substrates. Actually, the substitution of ruthenium by iridium has no additional effect so that all the observed peculiarities are caused by graphene layer. As for the response enhancement, $J_z \langle s_z \rangle$ in the right-hand part of Eq. 7 acts as an effective magnetic field to magnetize the magnetic impurities. At the same time, $J_z \langle S_z \rangle$ acts as the effective magnetic field to polarize the electron spin of TI. Obviously, such a double action of the exchange coupling leads to the enhancement of the magnetic response. When magnetic impurities form a continuous adlayer, additional enhancement should be expected due to the magnetic proximity effect (see one of the last publications [57] and references therein). Therefore, empirically confirmed graphene behaves as typical TI, which leads to a severe reconsideration of its physical properties discussed mainly without taking into account this important fact.

**4. Local Spins in Graphene Molecule Landscape**

Local spins of graphene, which are actively involved in the manifestation of its topological non-triviality discussed above, are associated with effectively unpaired $p_z$ electrons and are one of the most important characteristics of the UHF formalism applied to the graphene molecule open-shell ground state [3]. As seen in Table 1, bare graphene molecules are characterized by rather big total numbers of such electrons $N_D$ that constitute more than one third of the total number of the odd $p_z$ electrons. It means that the molecules are strongly radicalized thus exhibiting a large chemical activity. While the total number of effectively unpaired electrons is the quantitative measure of the activity of the whole molecule, or *molecular chemical susceptibility* (MCS) introduced in [58], their partitioning over molecule atoms describes the *atomic chemical susceptibility* (ACS) in terms of $N_{DA}$ due to which maps of their distribution over atoms present chemical portrait of the associated molecules [59]. Such maps of graphene molecules have a very particular, but therewith standard image which allows both disclosing the local spin distribution over atoms and considering chemical activity of the molecules at the quantitative level.

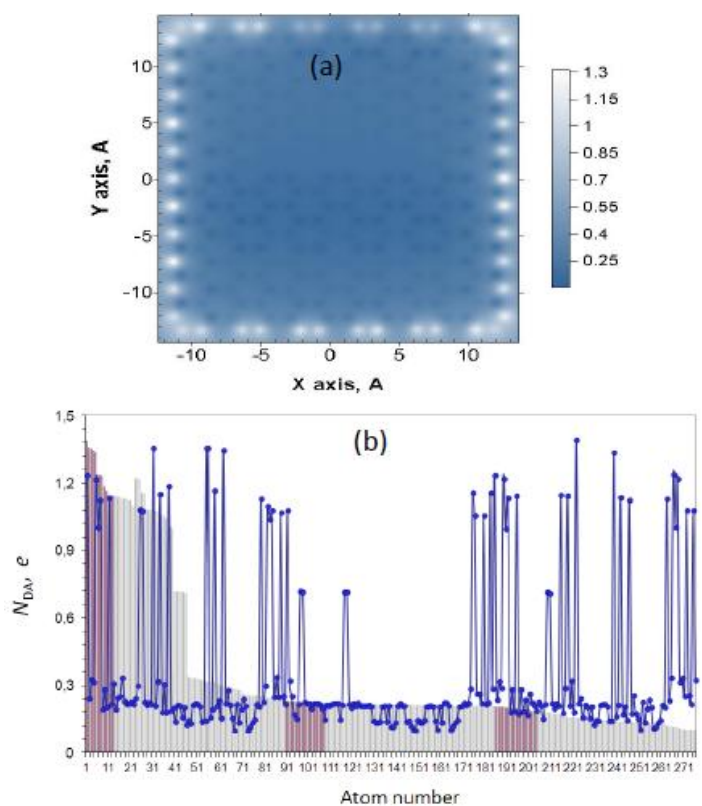

**Figure 10**. (a) Equilibrium structure and ACS $N_{DA}$ image map of the (11, 11) NGr molecule with bare edges. Scale bar matches $N_{DA}$ values in e. (b) $N_{DA}$ plotting from output file (curve with dots) and max→min $N_{DA}$ distribution (histogram). For the first 46 atoms: the histogram reveals first 22 zigzag edge atoms while next 24 bars (from 23 to 46) are related to armchair edge atoms. UHF AM1 calculations.

Figure 10 presents the $N_{DA}$ distribution (the ACS $N_{DA}$ image map below) over atoms of the (11, 11) NGr bare molecule. As seen in the figure, according to this parameter, the graphene molecule is definitely divided into two drastically different parts, namely, the circumference involving 46 edge atoms and internal honeycomb zone, or basal plane. Due to six-fold difference of the maximum $N_{DA}$ values in the two areas, the basal plane is practically invisible in Fig. 10a, while keeping a considerable $N_{DA}$ of ~0.2 e in average. The value rising over the average one occurs only for 40 atoms adjacent the molecule perimeter of edge atoms, for which $N_{DA}$ varies from 0.34 $e$ to 0.22 $e$. This atom fraction is clearly seen in the histogram in Fig. 10b at atom numbers from 47 to 86.

Presented in the figure is the chemical portrait of the bare (11, 11) NGr molecule. As seen from the histogram in Fig.10b, the chemical activity of the graphene molecule atoms greatly varied within both the circumference and basal plane, more significantly within the first one. In the histogram 46 edge atoms are divided into 22 and 24 atoms related to zigzag and armchair edges, respectively. For zigzag atoms $N_{DA}$ values fill the region 1.39-1.10 $e$, while the latter for armchair atoms is much wider and constitutes 1.22-0.71 $e$.

Qualitatively, the picture is typical to graphene sheets of any size. Modern fascinating experimental techniques allow confirming the above statement. Thus, as shown in [3, 7], the advanced atom-resolved AFM is able to fix local spins in pentacene molecule when using CO-terminated gold tip. Since this experiment provides monitoring of the molecule chemical activity, a close similarity of the AFM image and the relevant ACS $N_{DA}$ map occurred quite expected. Happily, such a justification is available now for (n, 3) graphene nanoribbon as well [60]. As in the case of pentacene, the scanning transmission electron microscope (STEM) image of the nanoribbons (Fig.11a) is in good consent with the calculated $N_{DA}$ map of the (15, 3) NGr molecule (Fig. 11b) thus, in particular, evidently justifying a peculiar two-zone character of the graphene pool of effectively unpaired electrons.

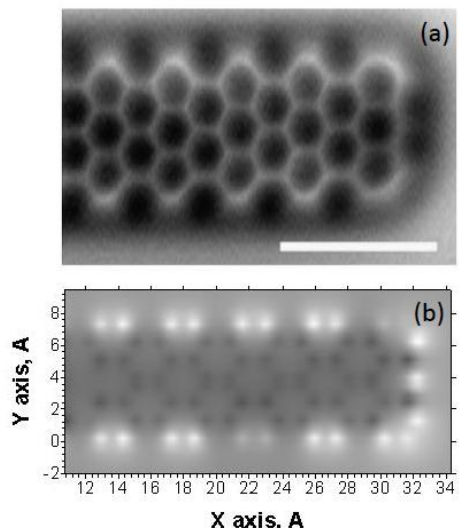

**Figure 11**. (a) Constant-height high-resolution AFM image of the zigzag end of a graphene nanoribbon obtained with a CO-terminated tip. White scale bar: 1 *nm*. Adapted from Ref. 60. (b) ACS $N_{DA}$ image map of the zigzag-end (15, 3) NGr molecule with bare edges. UHF AM1 calculations.

The variety of electron density of carbon atoms along edges of a graphene flake as well as perpendicular to them has been clearly demonstrated just recently by site-specific single-atom electron energy loss spectroscopy (EELS) by using annular dark field (ADF) mode of a low-voltage STEM [61]. Figure 12 discloses a highly informative picture related to the states of carbon atoms in the vicinity of zigzag and armchair edges. As seen in the figure, the site-dependent peculiarities are observed in the low-energy parts of the EELS spectra, which present K-edges of the carbon EELS signals, while the spectra above 290 *eV* are broad and less informative. All the low-energy spectra involve a characteristic EELS peak $P_b$ at 285 *eV* related to the excitation transformation of a core *s*-electron to an unoccupied $\sigma^*$ orbital. Additional peaks at 281 *eV* ($P_z$) and 282.8 *eV* ($P_a$) for the zigzag and armchair edge atoms, respectively, are caused by the *s* electron excitation to an unoccupied $p_z^*$ orbital and the change to the profile of the EELS is related to variations in the local density of states. The peaks are well pronounced for edge atoms (spectrum 1), significantly decrease in intensity for adjacent atoms (spectrum 2) and practically fully vanish for carbon atoms on the flake basal plane (spectra 3 and 4). Additionally, EELS spectra across the edge markedly vary for both zigzag and armchair atoms. As seen in Fig. 13a, the spectra of two neighboring zigzag atoms (8 and 10) differ so seriously that the peak $P_z$ is substituted by the peak $P_a$. The latter structure is conserved for the adjacent atom 9, albeit with changing in the intensity distribution between $P_a$ and $P_b$ peaks. EELS spectra in Fig.13b exhibit the difference in the behavior of the neighboring armchair atoms expressed in changing the $P_a/P_b$ intensities ratio.

The discussed spectral features are well consistent with the conclusion obtained from the above analysis of the $N_{DA}$ distribution in Fig. 10b. Thus, first, the chemical bonding of zigzag and armchair edge atoms is different, bigger in the latter case, which is consistent with lower chemical activity of the armchair edge atoms compared with the zigzag ones. Second, atoms of the adjacent-to-edge rows demonstrate a transient state between the edge and bulk one that well correlates with the activity of 40 adjacent atoms (from 47 to 86) in the histogram in Fig. 10b. Third, inside the region, perimeter of which is formed by adjacent atoms, the carbon atoms can be attributed to the basal-plane ones. Forth, electron density as well as ACS $N_{DA}$ of the edge, adjacent, and bulk atoms significantly varies thus demonstrating that the atom groups are not rigidly standardized and might be very sensitive to external actions due to which graphene molecules are very changeable.

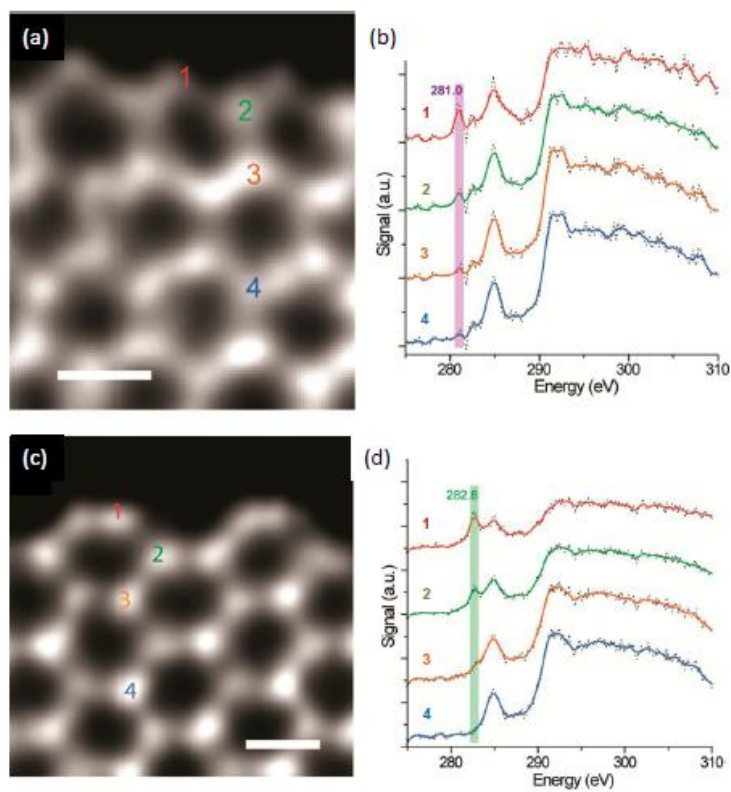

**Figure 12.** STEM: EELS mapping of graphene edges. (a) ADF-STEM image and (b) EELS of zigzag edge from the regions numbered in (a). (c) ADF-STEM image and (d) EELS of armchair edge from the regions numbered in (d). White scale bar: 2 Å. Numbers and their colors on ADF-STEM images and EELS spectra coincide. Adapted from Ref. 61.

The two-zone electron density image of bare graphene molecule is not new. The feature lays the foundation of a large number of theoretical-computational considerations concerning a particular role of edge atoms in graphene started in 1996 [62] and has been lasting until now (see a collection of papers [63-66] and references therein). The studied graphene objects were mainly pencil-made with a regular honeycomb structure described by standard C=C bonds of 1.42 Å in length and identical zigzag and armchair edge atoms. The obtained results concern the two-zone electronic structure and the attribution of the edge atoms peculiarity to expected particular magnetic behavior of graphene flakes and, particularly, nanoribbons. However, the latter expectations occurred quite illusive and as shown experimentally, magnetic behavior of graphene samples is not directly connected with peculiar features of their edge atoms. It is worthwhile to remain a skeptical comment of Roald Hoffmann concerning his "Small but strong lessons from chemistry to nanoscience" [67]: "There is a special problem that theory has with unterminated structures—ribbons cut off on the sides, polymers lacking ends. If passivation is not chosen as a strategy, then the radical lobes of the unterminated carbon atoms, or undercoordinated transition metals, will generate states that are roughly in the middle energetically, above filled levels, below empty levels in a typical molecule that has a substantial gap between filled and unfilled levels. If such levels—states, the physicists call them—are not identified as "intruder" states, not really real, but arising from the artifact of termination, they may be mistaken for real states in the band gap, important electronically. And if electrons are placed in them, there is no end to the trouble one can get into. These band gap states are, of course, the origin of the reactivity of the terminated but not passivated point, line, or plane. But they have little to do with the fundamental electronic structure of the material".

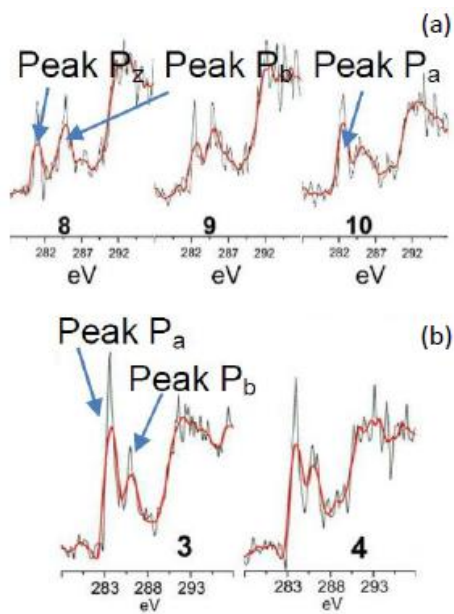

**Figure 13.** EELS spectra at the edges of graphene flake. (a) Neighboring zigzag atoms (atoms 8 and 9) and ajacent atom between them (atom 9). (b) Neighboring armchair atoms (atoms 3 and 4). P$_z$, P$_a$ and P$_b$ match peaks at $281 eV$, $282.8\ eV$, and $285 eV$, respectively. Adapted from Ref.61.

## 5. Introduction to Graphene Computational Spin Chemistry

The modern chemistry is strongly occupied by revealing reliable qualitative, better quantitative, descriptors aiming at pointing the consequence of chemical reaction. UHF formalism of open-shell molecules suggests unique quantitative descriptors MCS $N_D$ and ACS $N_{DA}$. For molecules with even number of electrons $N_{DA}$ is identical to the atom free valence [68]. Consequently, free valence of atom A, $V_A^{free}$, is defined as

$$V_A^{free} = N_A^{val} - \sum_{B \neq A} K_{AB} \tag{8}$$

Here $N_A^{val}$ is the number of valence electrons of atom A and $\sum_{B \neq A} K_{AB}$ presents a sum over the generalized bond index

$$K_{AB} = |P_{AB}|^2 + |D_{AB}|^2 \tag{9}$$

where the first term is the Wiberg bond index while the second term is determined by taking into account the spin density matrix. The $V_A^{free}$ distribution (curve with dots) alongside with the ACS $N_{DA}$ (histogram) for the (5, 5) NGr molecule is shown in Fig. 14. As seen in the figure, first steps of any chemical reaction occur at the molecule periphery. Since this reactivity area is largely spread in space, the formation of the first monoderivative does not inhibit the molecule reactivity so that the reaction will continue until the reaction ability is satisfied. This means that any chemical modification of graphene starts as polyderivatization of the pristine molecule at its circumference.

Excellent agreement of $N_{DA}$ and $V_A^{free}$ values shows that the former is actually a quantitative ACS measure and can serve a quantitative descriptor of the molecule target atoms, to which atom-atom contacts are the most desirable in addition reactions. Thus, the values distribution over molecule atoms forms a unique ACS $N_{DA}$ map, which opens a transparent methodology of a successive algorithmic computational synthesis of any graphene polyderivatives, just selecting the graphene core atom at each step by the largest $N_{DA}$ value. A successive use of this methodology was shown on examples of hydrogenation [69] and oxidation [70] of the (5, 5) NGr molecule.

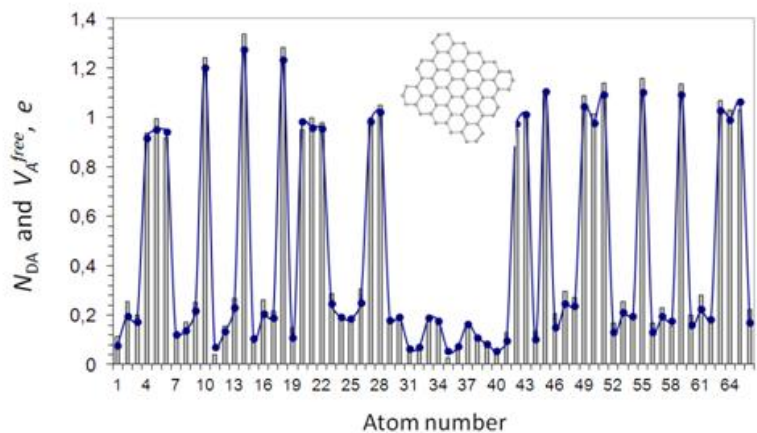

**Figure 14.** ACS $N_{DA}$ (histogram) and free valence $V_A^{free}$ (curve with dots) distributions over atoms of the (5, 5) NGr molecule. Inset: equilibrium structure of the (5, 5) NGr molecule. UHF AM1 calculations.

As turned out, already the first addition of any reactant, or modifier, to the edge atom of graphene molecule, chosen by the highest ACS, causes a considerable changing in the pristine ACS $N_{DA}$ image map thus allowing the exhibition of the second edge atom with the highest ACS to proceed with the chemical modification and so forth. This behavior is common to graphene molecules of any size and shape. In what follows, the behavior feature will be demonstrated on the example of the (5, 5) NGr molecule that was chosen to simplify the further presentation. Figure 15 presents a set of (5, 5) NGr polyhydrides and polyoxides obtained in the course of the first stage of the relevant per step reactions that concerns framing of the bare molecule. Two important conclusions follow from the figure. First, in spite of seemingly local change of the molecule structure caused by the addition, the second target carbon atoms does not correspond to the atom that is the second one of the highest activity in the $N_{DA}$ list of the pristine molecule. Second, this atom position as well as the sequence of steps varies depending on the chemical nature of the addends. Both two features are the result of the redistribution of C=C bond lengths over the molecule thus revealing collective action of its unpaired electrons and/or local spins.

**6. Comments on Converting Graphene from Semimetal to Semiconductor**

Despite numerous extraordinary properties and huge potential for various applications, one of the greatest challenges in using graphene as an electronic material is the lack of a sizable bandgap. Graphene is intrinsically a zero-gap semiconductor, or not a QSHI but a semimetal in the view of most. The gap absence significantly limits the use of graphene in many applications where semiconducting materials with a suitable bandgap are required. Researchers have been searching for effective ways to produce semiconducting graphene and have developed various methods to generate a bandgap in graphene. Despite extensive investigation in the laboratory, the production of semiconducting graphene is still facing many challenges. A detailed description of problems on the way as well as suggestions for their resolving is given in review [71]. Let us look at the problems from the viewpoint of obvious 'underwater stones' provided by the common properties of the graphene chemistry.

When categorizing methods to produce semiconducting graphene, three groups were classified: (1) morphological patterning of graphene sheets into nanoribbons, nanomeshes, or quantum dots to induce quantum confinement and edge effects; (2) chemical modification, doping, or surface functionalization of graphene to intentionally interrupt the connectivity of the π electron network; and (3) other methods, e.g., use of two graphene layers arranged in Bernal stacking (or AB stacking) to break the lattice symmetry, and applying mechanical deformation or strain to graphene.

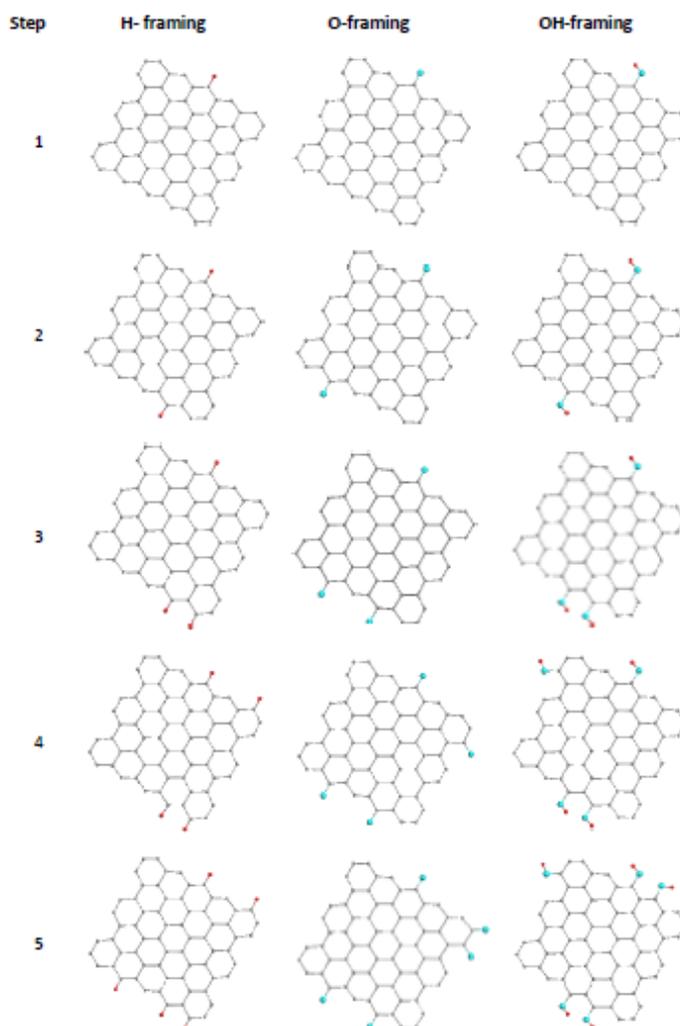

**Figure 15.** Equilibrium structures of the (5,5) NGr polyhydrides and polyoxides related to $1^{st}$, $2^{nd}$, $3^{rd}$, $4^{th}$, and $5^{th}$ steps obtained in the course of the relevant stepwise reactions. Gray, blue, and red balls mark carbon, oxygen, and hydrogen atoms, respectively. UHF AM1 calculations.

Following the scheme proposed in [71] and shown in Fig. 16, techniques of the first group meet problems concerning the basic edge property of the graphene molecule that is obviously a dangling-bond effect. Actually, cutting graphene sheets into nanoribbons increases the number of dangling bonds and, consequently, the number of unpaired electrons $N_D$ thus enhancing the ribbon radical properties. In its turn, the extra radicalization greatly promotes various chemical reactions in the ribbon circumference leading to significant and even sometimes drastic reconstruction of the pristine graphene structure. Inserting nanomeshes results in the same effect due to highly active periphery of the formed holes. Deposition of nanosize quantum dots highly disturbs the graphene substrate changing C=C bond length distribution and thus causing the $N_D$ growth if even not contributing by their own unpaired electrons. Therefore, cutting and drilling create a big 'edges problem' and do not seem to be proper technologies for the wished transformation of the graphene electronic system.

Chemical modification of graphene is not only a subject of interesting chemistry but has been repeatedly suggested as an efficient tool for the semimetal-semiconductor transferring needed for high-performance electronics [71]. It should be noted that the suggestions are based on results of computational studies that concern pencil-drawn pictures of graphene fragments including those or other chemical modifiers artificially spread over graphene sheets (see, for example, Refs. 72 and 73). These and many other virtual structures, regularly distributed in space by applying periodic boundary conditions, exhibit electronic properties that are so badly needed for the application. However, the empirical reality is much less promising since so far none of regularly chemically modified graphene structure has been obtained. And collective behavior of graphene unpaired electrons, protesting against any response localization, is the main reason for the failure.

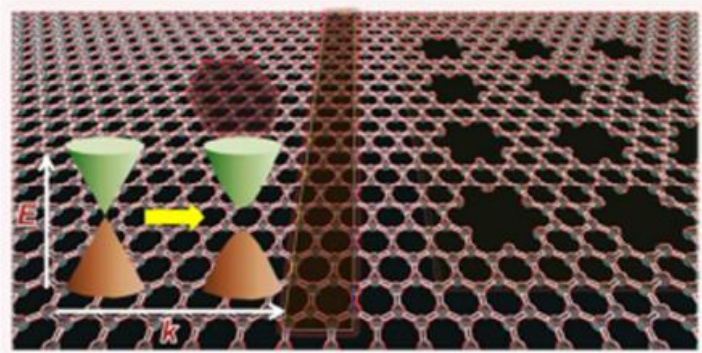

**Figure 16**. A schematic view of morphological patterning of a graphene sheet. Adapted from Ref. 71.

The wished regular structures of chemically modified graphene are related to the graphene polyderivatives that are formed with the participation of carbon atoms on the basal plane. However, as was shown earlier, reactions at the circumference precede those at the basal plane. Moreover, the latter cannot begin until the former are completed. In the predominant majority of the studied cases, the completion of the circumference reactions means the completion of framing of the studied molecules. A thorough study of the circumference reactions has disclosed a very exciting feature: the framing of graphene molecules promotes the molecule cracking. Figure 17 presents a set of ACS $N_{DA}$ maps related to mono-hydrogen terminated ($H_1$-terminated below) NGr molecules of different size. The ACS maps of all the pristine molecules are of identical pattern characteristic for the (11, 11) NGr molecule shown in Fig. 10 just scaled according to the molecule size. As seen in the figure, the ACS maps of $H_1$-terminated polyderivatives show a peculiar two-part division related to (15, 12) (3.275 x 2.957 $nm^2$) and (11, 11) (2.698 x 2.404 $nm^2$) NGr molecules in contrast to the maps of (9, 9) (1.994 x 2.214 $nm^2$), (7, 7) (1.574 x 1.721 $nm^2$), and (5, 5) (1.121 x 1.219 $nm^2$) NGr molecules. Apparently, the finding demonstrates the ability of graphene molecules to be divided when their linear size exceeds 1-2 $nm$. The cracking of pristine graphene sheets in the course of chemical reaction, particularly, during oxidation, was repeatedly observed. A peculiar size effect was studied for graphene oxidation [74] and fluorination [75]. During 900 sec of continuous oxidation, micrometer graphene sheets were transformed into ~1 $nm$ pieces of graphene oxide. Obviously, the tempo of cracking should depend on particular reaction conditions, including principal and service reactants, solvents, temperature, and so forth (see [76, 77]). Probably, in some cases, cracking can be avoided. Apparently, this may depend on particular conditions of the inhibition of the edge atoms reactivity. However, its ability caused by the inner essence of the electron correlation is an imminent threat to the stability and integrity of the final product.

In some cases, the cracking is not observed when graphene samples present membranes fixed over their perimeter on solid substrates. Therewith, the reactivity of circumference atoms is inhibited and the basal plane is the main battlefield for the chemical modification. Still, as in the case of circumference reactions considered earlier, the highest ACS retains its role as a pointer of the target carbon atoms for the subsequent reaction steps. However, the situation is much more complicated from the structural aspect viewpoint. Addition of any modifier to the carbon atom on the basal plane is accompanied by the $sp^2 \rightarrow sp^3$ transformation of the valence electrons hybridization so that for regularly packed chemical derivatives, the benzenoid skeleton of pristine graphene should be substituted by the cyclohexanoid one related to the formed polyderivatives. When benzene molecules and, subsequently, benzenoid units are monomorphic, cyclohexanes, and thus cyclohexanoid units, are highly heteromorphic. Not very big difference in the conformers free energy allows for coexisting cyclohexanoids of different structure thus making the formation of a regular structure a rare event. Actually, the regular crystalline-like structure of a graphene polyhydride, known as graphane, was obtained experimentally under particular conditions only when hydrogenating fixed graphene membranes accessible to hydrogen atoms from both sides [78]. In the same experiment, fixed membranes accessible to hydrogen atoms from one side showed irregular amorphous-like structure. The empirical findings were supported by computations based on the consideration of stepwise hydrogenation of fixed and free standing membranes accessible to hydrogen atom from either two or one side [69].

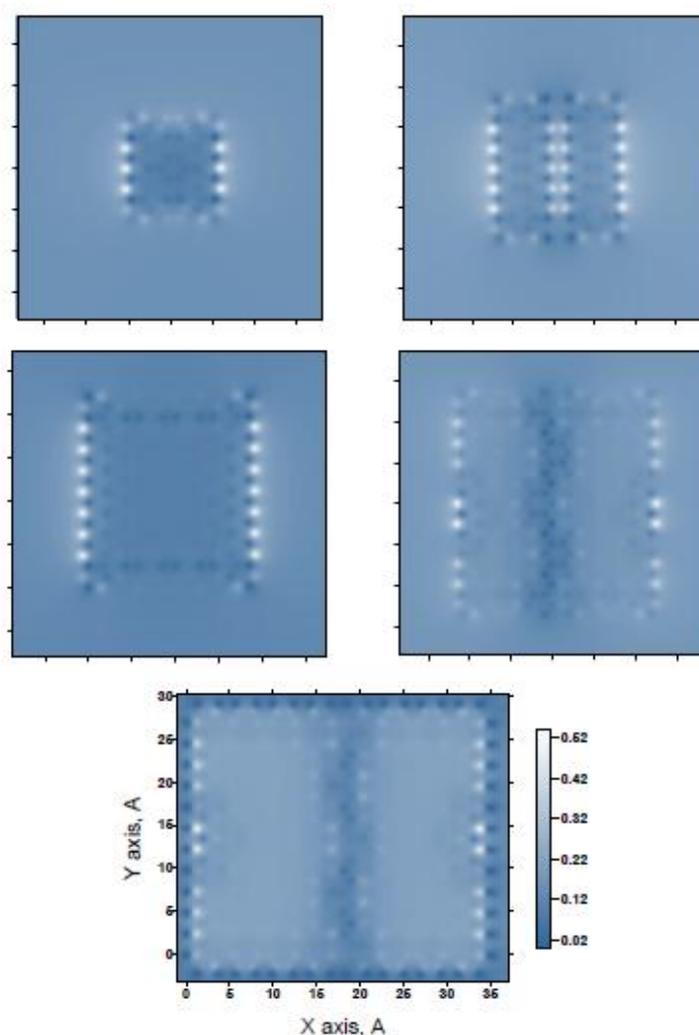

**Figure 17.** ACS $N_{DA}$ image maps over atoms of the (5, 5), (7, 7), (9, 9), (11, 11), and (15, 12) NGr molecules with $H_1$-terminated edges. All the images are given in the same space and $N_{DA}$ scales shown on the bottom. UHF AM1 calculations.

As shown above, it is possible to proceed with chemical modification of graphene within the basal plane only after a complete inhibition of high chemical activity of atoms at the circumference. Despite the $N_{DA}$ values within the area are much less than at a bare circumference, as seen in Fig. 10b, they still constitute ~0.3-0.1 $e$ that is quite enough to maintain active chemical modification. However, the reality drastically differs from the wished chemical pattering of graphene sheets whose virtual image present the final product of the pattering as regular carpets similar to flowerbeds of the French parks. The reality is more severe and closer to designs characteristic of the English parks. The matter is that the collective of unpaired electrons, which strictly controls the chemical process at each step, has no means by which to predict the modifier deposition sites many steps forward. And it is clear why. Each event of the modifier deposition causes an unavoidable structure deformation due to local $sp^2 \rightarrow sp^3$ transformation in the place of its contact with graphene. The relaxation of the deformation, as was seen in Fig. 15, extends over a large area, which, in turn, is accompanied by the redistribution of C=C bond lengths. Trying to construct a pattering, it is impossible, while not making calculations, to guess at what exactly carbon atom will concentrate the maximum reactivity, highlighting the latter as a target atom to the next deposition. Therefore, even two simultaneous depositions cannot be predicted, not to mention such complex as quantum dots or nanoribbons. That is why a wished regular chemical pattering of graphene basal plane exists only in virtuality. The real situation was studied in detail in the case of graphene hydrogenation [79], exhibiting the gradual filling of the basal plane with hydrogen at random. Final products of the addition reactions on basal planes of graphene strongly depend on the addends in use. None of the regular motives was observed in all the cases in the course of stepwise reactions.

As for use of graphene bi- and multilayers and applying mechanical deformation or strain to graphene, each of the technique has its own limitation since again any structural changing affects the pool of effectively unpaired electrons (local spins) whose reaction is complex and nonlocal as well as practically no predictable.

## 7. Conclusion

Graphene is deeply spin-rooted species, starting with quasi-relativistic description of the electronic state of graphene hexagonal honeycomb structure and finishing with topological non-triviality of graphene crystals and molecules. The latter is convincingly supported with the magnetic behavior of graphene molecules and is the consequence of local spin emergence over carbon atoms subordinated to zero spin density of the species. Local spins form the ground for a computational spin chemistry of graphene and free from the illusions associated with tuned morphological and chemical modification of graphene towards its converting from gapless QSHI to semiconductor.

## Acknowledgements

The author takes the opportunity to express her deep gratitude to I.S.Burmistrov and Yu.E.Lozovik for fruitful and stimulating discussions.

## References

1. *Graphene Science Handbook: 6-volume set* (2016). Eds. Aliofkhazraei, M., Ali, N., Miln, W.I., Ozkan C. S., Mitura, S. and Gervasoni, J. (CRC Press, Taylor and Francis Group, Boca Raton.
2. Löwdin, P-O. (1958). Correlation problem in many-electron quantum mechanics. 1. Review of different approaches and discussion of some current ideas, *Adv. Chem. Phys.,* **2**, pp. 209-322.
3. Sheka, E.F. (2014) The uniqueness of physical and chemical natures of graphene: Their coherence and conflicts, *Int. J. Quant. Chem.*, **114**, pp. 1079-1095.
4. Sheka, E.F. (2012) ) Computational strategy for graphene: Insight from odd electrons correlation, Int. J. Quant. Chem., 112, pp. 3076-3090.
5. Sheka, E.F. and Chernozatonskii, L.A. (2010) Chemical reactivity and magnetism of graphene, *Int. J. Quant. Chem.*, **110**, pp.1938-1946Sheka et al mechanics
6. Sheka, E.F., Popova, N.A., Popova, V.A., Nikitina, E.A., Shaymardanova, L.Kh. (2011) Structure-sensitive mechanism of nanographene failure. *J. Exp. Theor. Phys.,* **112**, pp. 602-611.
7. Sheka, E. F. (2016). Spin effects of $sp^2$ nanocarbons in light of unrestricted Hartree-Fock approach and spin-orbit coupling theory, in *Advances in Quantum Methods and Applications in Chemistry, Physics, and Biology* (Tadjer, A., Brändas, E.J., Maruani, J., Delgado-Barrio, G., ed.) Progress in Theoretical Chemistry and Physics 31, Springer, Switzerland, pp. xxx-yyy.
8. Wallace, P.R. (1947). The band theory of graphite, *Phys. Rev.,* **71**, pp. 622-634.
9. Kane, C. L. and Mele, E. J. (2005) Quantum spin Hall effect in graphene, *Phys. Rev. Lett.,* **95**, 226801
10. Guzmàn-Verri, G.G. (2006). *Electronic Properties of Silicon-Based Nanostructures*. MS thesis, Wright State University, Dayton.

11. Slonczewski, J.C. and Weiss, P.R. (1957). Band structure of graphite, *Phys. Rev.,* **109,** pp. 272-279.
12. Katsnelson, M.I. (2007). Graphene: carbon in two dimensions, *Materials Today,* **10,** pp. 20-27.
13. Geim, A.K. and Novoselov, K.S. (2007). The rise of graphene, *Nat. Mat.,* **6,** pp. 183-191.
14. Kim, P. (2014). Graphene and relativistic quantum physics, *Matiere de Dirac*, Seminaire Poincare XVIII: pp. 1-21.
15. Hwang, C., Siegel, D.A., Mo, S.-K., Regan, W., Ismach, A., Zhang, Y., Zettl, A. and Lanzara, A. (2012). Fermi velocity engineering in graphene by substrate modification, *Sci. Rep.,* **2**, 590.
16. Kara, A., Enriquez, H., Seitsonen, A.P., Lew, Yan Voon, L.C., Vizzini, S., Aufray, B., Oughaddou, H. (2012). A review on silicene — New candidate for electronics, *Sur.f Sci. Rep.,* **67,** pp. 1−18.
17. Sheka, E.F. (2016) Silicene is a material phantom. Nanosyst. Phys. Chem. Math., **7**, xxx-yyy.
18. Zhang, R.-w., Ji, W.-x., Chang-wen Zhang, C.-w., Li, P. and Wang, P.-j. (2016) Prediction of flatness-driven quantum spin Hall effect in functionalized germanene and stanene. *Phys. Chem. Chem. Phys.* **18**, pp. 28134-28139.
19. Gomes, K.K., Mar, W., Ko, W., Guinea, F. and Manoharan, H.C. (2012). Designer Dirac fermions and topological phases in molecular graphene, *Nature,* **483,** pp. 307-311.
20. Tarruell, L., Greif, D., Uehlinger, T., Jotzu, G. and Esslinger, T. (2012). Creating, moving and merging Dirac points with a Fermi gas in a tunable honeycomb lattice, *Nature,* **483,** pp. 302-306.
21. Bhimanapati, G.R., Lin, Z., Meunier, V., Jung, Y., Cha, J., Das, S., Xiao, D., Son, Y., Strano, M.S., Cooper, V.R., Liang, L., Louie, S.G., Ringe, E., Zhou, W., Sumpter, B.G., Terrones, H., Xia, F., Wang, Y., Zhu, J., Akinwande, D., Alem, N., Schuller, J.A., Schaak, R.E., Terrones, M. and Robinson, J.A. (2015). Recent advances in two-dimensional materials beyond graphene, *ACS Nano*, **9**, pp. 11509–11539.
22. Xu, L.-C., Du, A. and Kou, L. (2016) Hydrogenated borophene as a stable two-dimensional Dirac material with an ultrahigh Fermi velocity. Phys. Chem. Chem. Phys., 18, 27284-27289.
23. Wang, C., Xia, Q., Nie, Y., Rahman, M. and Guo, G. (2016) Strain engineering band gap, effective mass and anisotropic Dirac-like cone in monolayer arsenene, *AIP Advances,* **6,** 035204.
24. Wang, A., Zhang, X. and Zhao, M (2014). Topological insulator states in a honeycomb lattice of s-triazines, *Nanoscale,* **6,** pp. 11157-11162.
25. Zhang, X., Wang, A. and Zhao, M. (2015). Spin-gapless semiconducting graphitic carbon nitrides: A theoretical design from first principles, *Carbon,* **84,** pp. 1-8.
26. Wei, L., Zhang, X. and Zhao, M. (2016). Spin-polarized Dirac cones and topological nontriviality in a metal-organic framework $Ni_2C_{24}S_6H_{12}$, *Phys. Chem. Chem. Phys.*, **18**, 8059-8064.
27. Zhang, H., Li, Y., Hou, J., Du, A. and Chen, Z. (2016) Dirac state in the FeB2 monolayer with graphene-like boron sheet. *Nano Lett.* 16, pp. 6124–6129.
28. Si, C., Jin, K.-H., Zhou, J., Sun, Z. and Liu, F. (2016) Large-gap quantum spin hall state in MXenes: d-Band topological order in a triangular lattice. Nano Lett. DOI: 10.1021/acs.nanolett.6b03118.
29. Naguib, M., Mochalin, V.N., Barsoum, M.W. and Gogotsi, Y. (2014) 25th Anniversary Article: MXenes: A new family of two-dimensional materials. Adv. Mat. 26, 992–1005.
30. Bandurin, D.A., Tyurnina, A.V., Yu, G.L., Mishchenko, A., Zólyomi, V., Morozov, S.V., Kumar, R.K., Gorbachev, R.V., Kudrynskyi, Z.R., Pezzini, S., Kovalyuk, Z.D., Zeitler, U., Novoselov, K.S., Patanè, A., Eaves, L., Grigorieva, I.V., Fal'ko, V.I., Geim, A.K. and Cao, Y. (2016) High electron mobility, quantum Hall effect and anomalous optical response in atomically thin InSe, *Nat. Nanotech.* DOI: 10.1038/NNANO.2016.242
31. Sheka, E.F. (2015). Stretching and breaking of chemical bonds, correlation of electrons, and radical properties of covalent species, *Adv. Quant. Chem.*, **70**, pp. 111-161.
32. Hohenadler, M. and Assaad, F.F. (2013) Correlation effects in two-dimensional topological insulators, *J. Phys.: Condens. Matter,* **25**, 143201 (31pp).
33. Mayorov, A. S., Elias, D. C., Mukhin, I. S., Morozov, S. V., Ponomarenko, L. A., Novoselov, K. S., Geim, A.K. and Gorbachev, R. V. (2012). How close can one approach the Dirac point in graphene experimentally? *Nano Letters*, *12*, pp. 4629-4634.
34. *Topological Insulators: Fundamentals and Perspectives* (2015) Eds. Ortmann, F., Roche, S., Valenzuela, S.O., Molenkamp, L.W. Wiley: Chichester.
35. Schüler, M., Rösner, M., Wehling, T.O., Lichtenstein, A.I. and Katsnelson, M.I. (2013) Optimal Hubbard models for materials with nonlocal Coulomb interactions: graphene, silicene and benzene, *Phys. Rev. Lett.*, **111**, 036601.
36. Bučinský, L., Malček, M., Biskupič, S., Jayatilaka, D., Büchel, G.E. and Arion, V.B. (2015). Spin contamination analogy, Kramers pairs symmetry and spin density representations at the 2-component unrestricted Hartree–Fock level of theory, *Comp. Theor. Chem.*, **1065**, pp. 27-41.
37. Yazyev, O.V. (2010). Emergence of magnetism in graphene materials and nanostructures. *Rep. Prog. Phys.,* **73,** 05650130.
38. Esquinazi, P., Spemann, D., Hohne, R., Setzer, A., Han, K. H. and Butz, T. (2003). Induced magnetic ordering by proton irradiation in graphite, *Phys. Rev. Lett.,* **91,** 227201.

39. Sepioni, M., Nair, R.R., Rablen, S., Narayanan, J., Tuna, F., Winpenny, R., Geim, A.K. and Grigorieva, I.V. (2010). Limits on intrinsic magnetism in graphene, *Phys. Rev. Lett.,* **105,** 207205.
40. Nair, R.R., Sepioni, M., Tsai, I.-L., Lehtinen, O., Keinonen, J., Krasheninnikov, A.V., Thomson, T., Geim, A.K. and Grigorieva, I.V. (2012). Spin-half paramagnetism in graphene induced by point defects, *Nat. Phys.,* **8,** pp. 199-202.
41. Eng, A.Y.S., Poh, H.L., Sanek, F., Marysko, M., Matejkova, S., Sofer, Z. and Pumera, M. (2013). Searching for magnetism in hydrogenated graphene: Using highly hydrogenated graphene prepared via birch reduction of graphite oxides, *ACS Nano,* **7,** pp. 5930–5939.
42. Nair, R.R., Tsai, I.-L., Sepioni, M., Lehtinen, O., Keinonen, J., Krasheninnikov, A.V., Castro Neto, A.H., Katsnelson, M.I., Geim, A.K. and Grigorieva, I.V. (2013). Dual origin of defect magnetism in graphene and its reversible switching by molecular doping, *Nat. Commn.,* **4,** 2010.
43. Tada, K., Haruyama, J., Yang, H. X., Chshiev, M., Matsui, T. and Fukuyama, H. (2011). Ferromagnetism in hydrogenated graphene nanopore arrays, *Phys. Rev. Lett.,* **107,** 217203.
44. Ning, G., Xu, C., Hao, L., Kazakova, O., Fan, Z., Wang, H., Wang, K., Gao, J., Qian, W. and Wei, F. (2013). Ferromagnetism in nanomesh graphene, *Carbon,* **51,** pp. 390-396.
45. Van Fleck, J.H. (1932). *The Theory of Electric and Magnetic Susceptibilities*. Oxford.
46. Adamo, C., Barone, V., Bencini, A., Broer, R., Filatov, M., Harrison, N.M., Illas, F., Malrieu, J.P. and Moreira, I. de P.R. (2006). Comment on "About the calculation of exchange coupling constants using density-functional theory: The role of the self-interaction error" [J. Chem. Phys.123, 164110 (2005)], *Journ. Chem.Phys.,* **124,** 107101.
47. Noodleman, L. (1981). Valence bond description of antiferromagnetic coupling in transition metal dimmers, *J. Chem. Phys.,* **74,** pp. 5737-5742.
48. Kahn, O. (1993). *Molecular Magnetism*. VCH, New York.
49. Zayets, V.A. (1990). *CLUSTER-Z1: Quantum-Chemical Software for Calculations in the s,p-Basis,* Institute of Surface Chemistry Nat Ac Sci of Ukraine: Kiev
50. Gao, X., Zhou, Z., Zhao, Y., Nagase, S., Zhang, S.B. and Chen, Z. J. (2008). Comparative study of carbon and BN nanographenes: Ground electronic states and energy gap engineering, *Phys. Chem. A,* **112,** 12677.
51. Enoki, T. and Kobayashi, Y. (2005). Magnetic nanographite: an approach to molecular magnetism, *J. Mat. Chem.,* **15,** 3999.
52. Sheka, E.F., Zayets, V.A. and Ginzburg, I.Ya. (2006). Nanostructural magnetism of polymeric fullerene crystals, *J. Exp. Theor. Phys.,* **103,** pp. 728-739.
53. Nai, C.T., Xu, H., Tan, S.J.R., and Loh, K.P. (2016). Analyzing Dirac cone and phonon dispersion in highly oriented nanocrystalline graphene, *ACS Nano,* **10,** pp. 1681–1689.
54. Liu,Q., Liu, C.-X., Xu, C., Qi, X.-L. and Zhang, S.-C. (2009). Magnetic impurities on the surface of a topological insulator, *Phys. Rev. Lett.*, **102**, 156603.
55. Checkelsky, J.G., Ye, J., Onose, Y., Iwasa, Y. and Tokura, Y. (2012). Dirac-fermion-mediated ferromagnetism in a topological insulator, *Nature Phys.*, **8,** pp. 729-733.
56. Dreiser, J., Pacchioni, G.E., Donati, F., Gragnaniello, L., Cavallin, A., Pedersen, K.S., Bendix, J., Delley, B., Pivetta, M., Rusponi, S. and Brune, H. (2016). Out-of-plane alignment of Er(trensal) easy magnetization axes using graphene, *ACS Nano,* 10, pp. 2887–2892.
57. Katmis, F., Lauter, V., Nogueira, F.S., Assaf, B.A., Jamer, M.E., Wei, P., Satpati, B., Freeland, J.W., Eremin, I., Heiman, D., Jarillo-Herrero, P. and Moodera, J.S. (2016). A high-temperature ferromagnetic topological insulating phase by proximity coupling, *Nature*, **533**, pp. 513–516.
58. Sheka, E.F. (2007). Chemical susceptibility of fullerenes in view of Hartree-Fock approach, *Int. J. Quant. Chem.,* **107,** pp. 2803-2816.
59. Sheka, E.F. (2006). Chemical portrait of fullerenes, *J. Struct. Chem.,* **47,** pp. 593-599.
60. van der Lit, J., Boneschanscher, M.P., Vanmaekelbergh, D., Ijäs, M., Uppstu, A., Ervasti, M., Harju, A., Liljeroth, P. and Swart, I. (2013). Suppression of electron–vibron coupling in graphene nanoribbons contacted via a single atom, *Nat. Commn.,* **4,** 2023.
61. Warner, J.H., Lin, Y.-C., He, K., Koshino, M. and Suenaga, K.(2014). Atomic level spatial variations of energy states along graphene edges, *Nano Lett.,* **14,** pp. 6155−6159.
62. Nakada, K. and Fujita, M. (1996). Edge state in graphene ribbons: Nanometer size effect and edge shape dependence, *Phys. Rev. B,* **54,** pp. 17954-17961.
63. Barnard, A.S. and Snook, I.K. (2011). Modelling the role of size, edge structure and terminations on the electronic properties of graphene nano-flakes, *Modelling Simul. Mater. Sci. Eng.,* **19,** 054001.
64. Acik, M. and Chabal, Y.J. (2011). Nature of graphene edges: A review, *Jpn. J. Appl. Phys.,* **50,** 070101.
65. Mishra, P.C. and Yadav, A. (2012). Polycyclic aromatic hydrocarbons as finite size models of graphene and graphene nanoribbons: Enhanced electron density edge effect, *Chem. Phys.,* **402,** pp. 56–68.
66. Ang, L.S., Sulaiman, S. and Mohamed-Ibrahim, M.I. (2013). Effects of size on the structure and the electronic properties of graphene nanoribbons, *Monatsh. Chem.,* **144,** pp. 1271–1280.
67. Hoffmann, R. (2013). Small but strong lessons from chemistry for nanoscience, *Ang. Chem. Int. Ed.,* **52,** pp. 93-103.

68. Mayer, I. (1986). On bond orders and valences in the *ab initio* quantum chemical theory, *Int. J. Quant. Chem.,* **29,** pp. 73-84.
69. Sheka, E.F., Popova, N.A. (2012) Odd-electron molecular theory of the graphene hydrogenation. *J. Mol. Model.,* 18, pp. 3751–3768.
70. Sheka, E., Popova, N. (2013) Molecular theory of graphene oxide. *Phys. Chem. Chem. Phys.,* **15**, pp.13304-13332.
71. Lu, G., Yu, K., Wen, Z. and Chen, J. (2013). Semiconducting graphene: converting graphene from semimetal to semiconductor, *Nanoscale,* **5,** pp. 1353-1367.
72. Chernozatonski, L. A., Sorokin, P. B., Belova, E. E. and Brüning, J. (2007). Superlattices consisting of 'lines' of adsorbed hydrogen atom pairs on graphene, *JEPT Lett.,* **85,** pp. 77-81.
73. Lu, N., Huang, Y., Li, H-b., Li, Z. and Yang, J. (2010). First principles nuclear magnetic resonance signatures of graphene oxide, *J. Chem. Phys.,* **133,** 034502.
74. Pan, S. and Aksay, I.A. (2011). Factors controlling the size of graphene oxide sheets produced via the graphite oxide route, *ACS Nano,* **5,** pp. 4073-4083.
75. Nebogatikova, N.A., Antonova, I.V., Prinz, V.Ya., Kurkina, I.I., Vdovin, V.I., Aleksandrov, G.N., Timofeev, V.B., Smagulova, S.A., Zakirova, E.R. and Kesler, V.G. (2015). Fluorinated graphene dielectric films obtained from functionalized graphene suspension: preparation and properties, *Phys. Chem. Chem. Phys.,* **17,** pp. 13257-13266.
76. Wang, X., Bai, H. and Shi, G. (2011). Size fractionation of graphene oxide sheets by pH-assisted selective sedimentation, *J. Am. Chem. Soc.,***133,** pp. 6338–6342.
77. Kang, J.H., Kim, T., Choi, J., Park, J., Kim, Y.S., Chang, M.S., Jung, H., Park, K., Yang, S.J. and Park, C.R. (2016). The hidden second oxidation step of Hummers method, *Chem. Mat.,* **28**, pp. 756–764.
78. Elias, D.C., Nair, R.R., Mohiuddin, T.M.G., Morozov, S.V., Blake, P., Halsall, M.P., Ferrari, A.C., Boukhvalov, D.W., Katsnelson, M.I., Geim, A.K. and Novoselov, K.S. (2009). Control of graphene's properties by reversible hydrogenation: evidence of graphane, *Science,* **323,** pp. 610-613.
79. Balog, R., Jorgensen, B., Nilsson, L., Andersen, M., Rienks, E., Bianchi, M., Fanetti, M., Lægsgaard, E., Baraldi, A., Lizzit, S., Sljivancanin, Z., Besenbacher, F., Hammer, B., Pedersen, T.G., Hofmann, P. and Hornekær, L. (2010). Bandgap opening in graphene induced by patterned hydrogen adsorption, *Nature Mat.,* **9,** pp. 315-319.